\documentclass[a4paper,10pt]{article}
\usepackage[utf8]{inputenc}
\usepackage{authblk}

\usepackage{graphicx}
\usepackage{epstopdf}
\usepackage{todonotes}
\usepackage{lineno,hyperref}
\usepackage{upgreek}
\usepackage{units}
\usepackage{multirow}
\usepackage{array}
\usepackage{amssymb}	
\usepackage{amsmath}	
\usepackage{textcomp}
\usepackage{pdfpages}

\usepackage{caption}
\usepackage{subcaption}

\title{Measuring the coefficient of moisture expansion of Hysol 9396 loaded with boron nitride powder}

\author[1,2]{Luise Poley}
\author[3]{Tim Jones}

\affil[1]{Department of Physics, Simon Fraser University, University Drive W, Burnaby, Canada}
\affil[2]{TRIUMF, Wesbrook Mall, Vancouver, Canada}
\affil[3]{Department of Physics, University of Liverpool, Oxford Street, Liverpool, United Kingdom}

\begin{document}

\maketitle

\begin{abstract}

The future ATLAS ITk strip tracker~\cite{TDRs} will consist of 17,000 silicon strip detector modules mounted on support structures called cores. Cores are assembled from a number of components, among others carbon fibre facings, honeycomb structure and carbon foam surrounding titanium tubes used for cooling, using a two-component epoxy (Hysol 9396) loaded with boron nitride for good thermal conductivity. The adhesive constitutes about \unit[20]{\%} of a core's weight.

During operation, the detector is cooled down to \unit[-40]{$^{\circ}$C} using bi-phase carbon dioxide and flushed with dry gas to prevent condensation. The effect of this temperature change has been simulated to study the impact of Coefficient of Thermal Expansion (CTE) mismatches between different materials and investigate resulting deformations and misalignment.

In addition to the shrinking of an adhesive during cooling, which can be estimated well using its known CTE, flushing the detector volume with dry gas removes the moisture contained in the adhesive, leading to an additional shrinking. In order to estimate the impact of shrinking during drying, the Coefficient of Moisture Expansion (CME) of Hysol samples with different contents of boron nitride as well as their overall moisture absorption were measured and their extent compared to the contraction associated with cooling.

\end{abstract}

\section{Introduction}

The ITk strip tracker~\cite{TDRs} for the ATLAS~\cite{ATLAS} detector will be comprised of silicon strip sensor modules which are glued directly onto carbon fibre support structures, which provide mechanical support and cooling through embedded titanium pipes. Support structures are assembled from carbon fibre face sheets with co-cured bus tapes, closeouts, carbon fibre honeycomb structures and titanium pipes embedded in carbon foam. The parts are connected using Hysol~9396, a two-component epoxy, filled with boron nitride powder for thermal conductivity, which makes up about \unit[20]{\%} of the support structure (by weight).

Extensive studies have been performed to study the thermal behaviour of the assembled support structure and simulate its behaviour after cooling: during operation, support structures will be cooled down from room temperature to \unit[-40]{$^{\circ}$C}, i.e. by about \unit[-60]{$^{\circ}$C}, resulting in deformations due to CTE mismatches of the involved components.

During operation, the detector is additionally going to be flushed with dry gas to prevent humidity condensation within the detector volume. Adhesives are known to expand or contract with increasing/decreasing humidity content, however the exact coefficient of moisture expansion (CME) for the adhesive in use was unknown.

Measurements were therefore conducted to investigate the CME of Hysol 9396 with boron nitride filling as planned to be used in the ATLAS strip tracker.

\section{Samples under investigation}

All samples under investigation were prepared using Hysol 9396, an epoxy adhesive mixed from two components in a ratio of 100:30 (resin:hardener). Additionally, in order to study the dependence of the CME on the filling content, samples with different filling percentages of boron nitride powder were prepared.

Samples were prepared by mixing a minimum of \unit[10]{g} of Hysol 9396 in a glue mixer (using a glue mixer at \unit[2,500]{u/min} for \unit[2.5]{min}). Depending on the intended content, a measured amount of boron nitride powder was added and mixed using an automated glue mixer.

\begin{figure}[htp]
\centering
\includegraphics[width=0.8\linewidth]{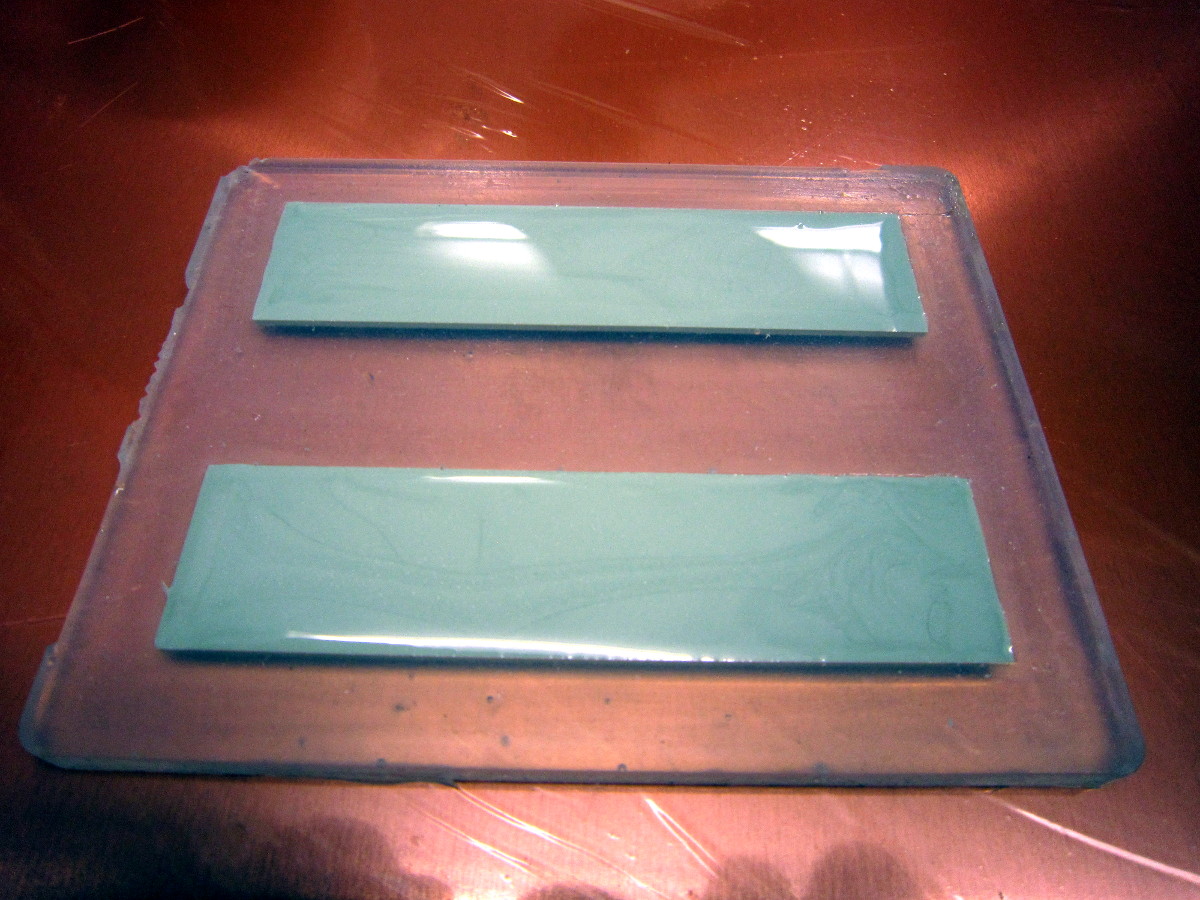}
\caption{Silicone mold with liquid Hysol 9396 (with {\unit[10]{\%}} boron nitride content) used to make samples with defined sizes ({\unit[$0.2-0.3 \times 2 \times 8$]{cm$^3$}})}
\label{fig:mold}
\end{figure}

A silicon mold was used to produce pairs of glue samples with defined dimensions of \unit[$2 \times 8$]{cm$^2$} and thicknesses of \unit[2-3]{mm} to maximise surface area and, thereby, moisture absorption (see figure~\ref{fig:mold}).

\begin{figure}[htp]
\centering
\includegraphics[width=0.8\linewidth]{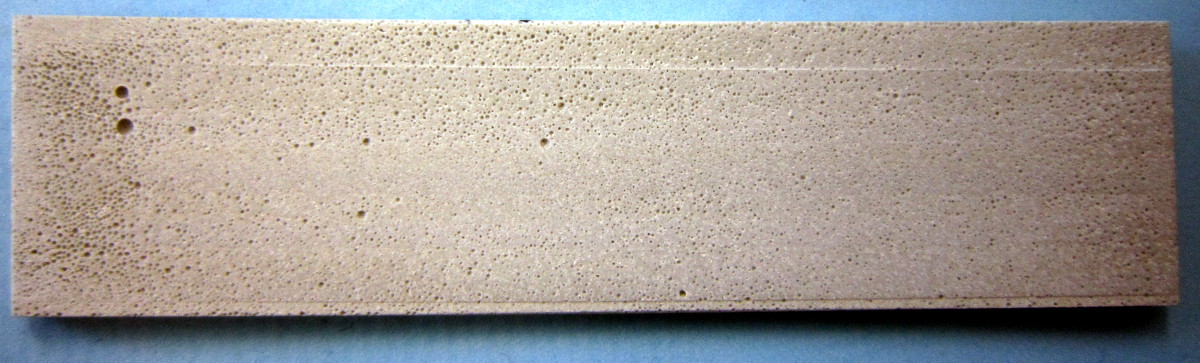}
\caption{Hysol 9396 sample with {\unit[20]{\%}} boron nitride powder content (by weight), milled open after oven curing: due to the fast curing of the sample at elevated temperatures air bubbles are trapped in the sample.}
\label{fig:bubbles}
\end{figure}

Since adding boron nitride powder introduced air bubbles in the glue mixture, which slowly escaped over time, two different approaches to curing were taken. Samples with \unit[0, 10 and 20]{\%} boron nitride content (by weight) were cured at elevated temperatures (\unit[60]{$^{\circ}$C} for \unit[1]{hour}), which led to faster curing and therefore more air being trapped within the sample (see figure~\ref{fig:bubbles}). All
other samples (with \unit[5, 15, 25 and 30]{\%} boron nitride content), were
cured at room temperature.

\begin{figure}
\begin{subfigure}{.47\textwidth}
  \centering
  \includegraphics[width=\linewidth]{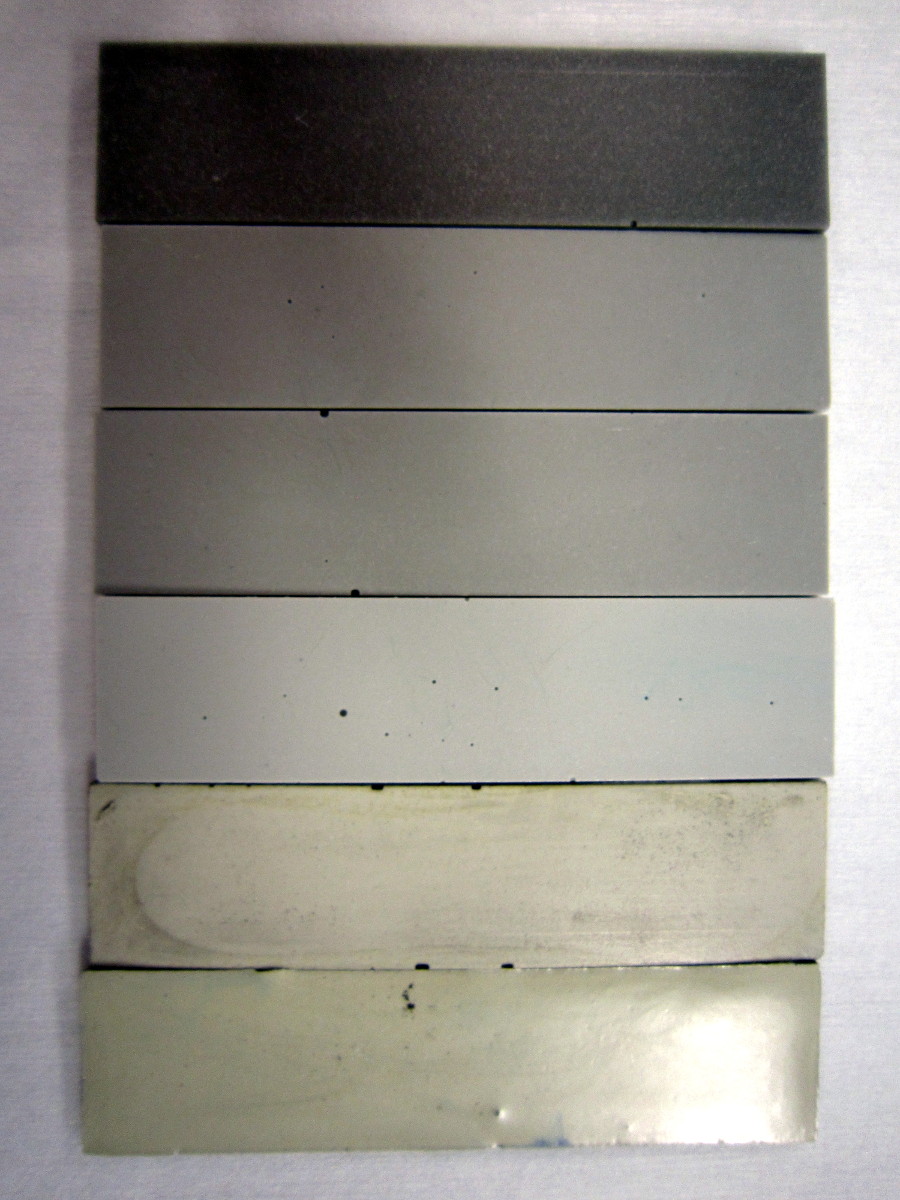}
  \caption{Selection of six glue samples after curing}
  \label{fig:milling1}
\end{subfigure}
  \begin{subfigure}{.04\textwidth}
\hfill
\end{subfigure}
 \begin{subfigure}{.47\textwidth}
  \centering
  \includegraphics[width=\linewidth]{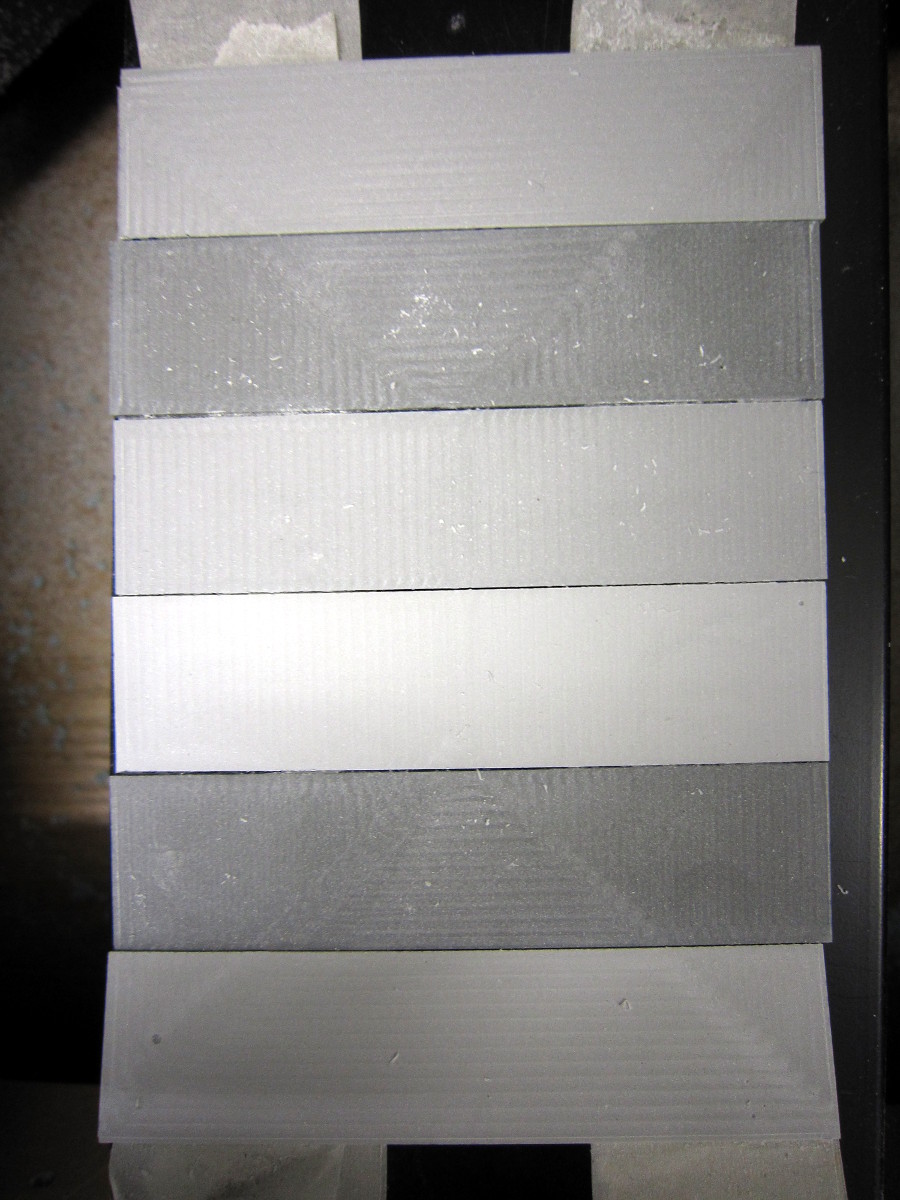}
  \caption{Selection of six glue samples after initial, coarse milling step}
  \label{fig:milling2}
\end{subfigure}
\caption{Selection of glue samples before and after milling the top surface off the samples. Several samples were machined at the same time to achieve uniform sample thicknesses.}
\label{fig:milling}
\end{figure}

Samples cured at both elevated temperatures and room temperature were found to have uneven surfaces after curing. CME measurements of these samples required measurements of their expansion due to moisture absorption, which were conducted using an image correlation system (see section~\ref{sec:imaging}), therefore sample surfaces were milled to provide even surfaces for imaging (see figures~\ref{fig:milling1} and~\ref{fig:milling2}). For each boron nitride content, one sample was milled down with the top surface up, and one sample was milled with the bottom surface up for comparison.
Afterwards, any marks left in samples surfaces by the milling process were sanded down to achieve a smooth sample surface.

\begin{figure}[htp]
\centering
\includegraphics[width=0.8\linewidth]{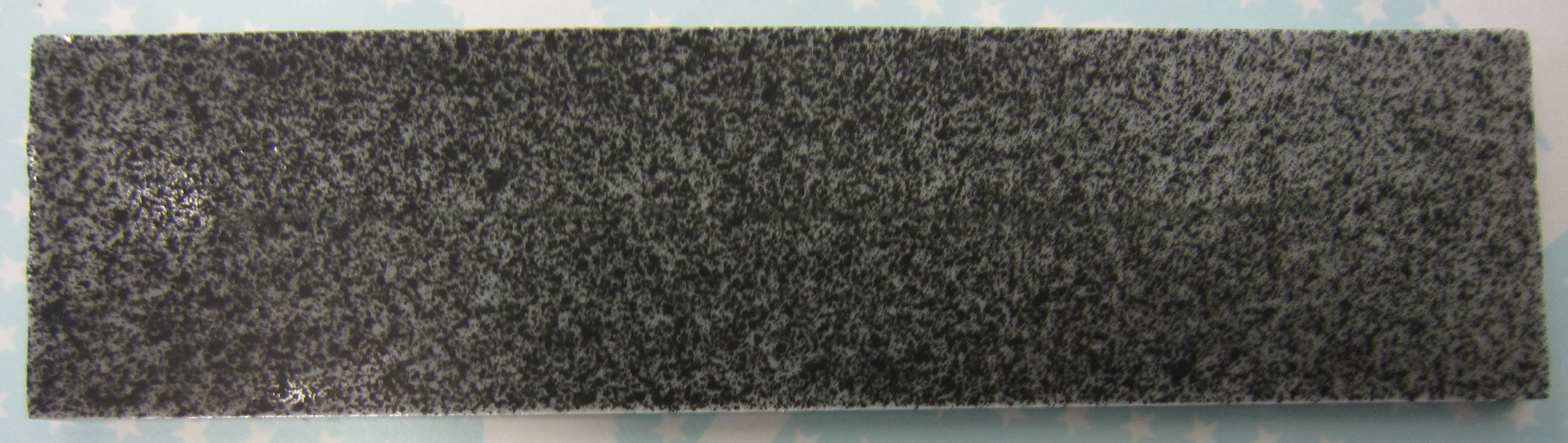}
\caption{Top surface of a Hysol sample (with a boron nitride content of {\unit[20]{\%}}), with a spray painted speckle pattern on the sample surface: high contrast, random speckle patterns were applied to the samples using black or white spray paint.}
\label{fig:speckles}
\end{figure}

Following these surface treatments, the foreseen use of an image correlation system required the application of a small-scale, high-contrast and random pattern~\cite{speckles}, see figure~\ref{fig:speckles}. In accordance with good practice recommendations~\cite{IMCOR}, spray paint was used to apply a pattern of small speckles that was
\begin{itemize}
\item matt (to minimise reflections)
\item resulted in similar numbers of light and dark pixels
\item added a layer of material that was thin compared to the overall sample thickness (about \unit[20]{$\upmu$m} of paint compared to a sample thickness of \unit[2]{mm})
\end{itemize}

\section{Moisture absorption}
\label{sec:absorb}

In preparation of CME measurements, all samples were baked out in order to determine their dry weight: all samples were stored in an oven flushed with dry air and baked at a temperature of \unit[30]{$^{\circ}$C}. They were weighed daily over a period of up to four weeks (depending on the sample), to find their dry weight (where their weight did not decrease further over a period of several days). The thereby determined dry weight was used for calculations of absorbed water content.

Measurements were performed with fully dried samples, since dry samples absorbed moisture faster and were therefore better suited to study the sample expansion over a larger range of water content.

\begin{figure}[htp]
\centering
\includegraphics[width=0.8\linewidth]{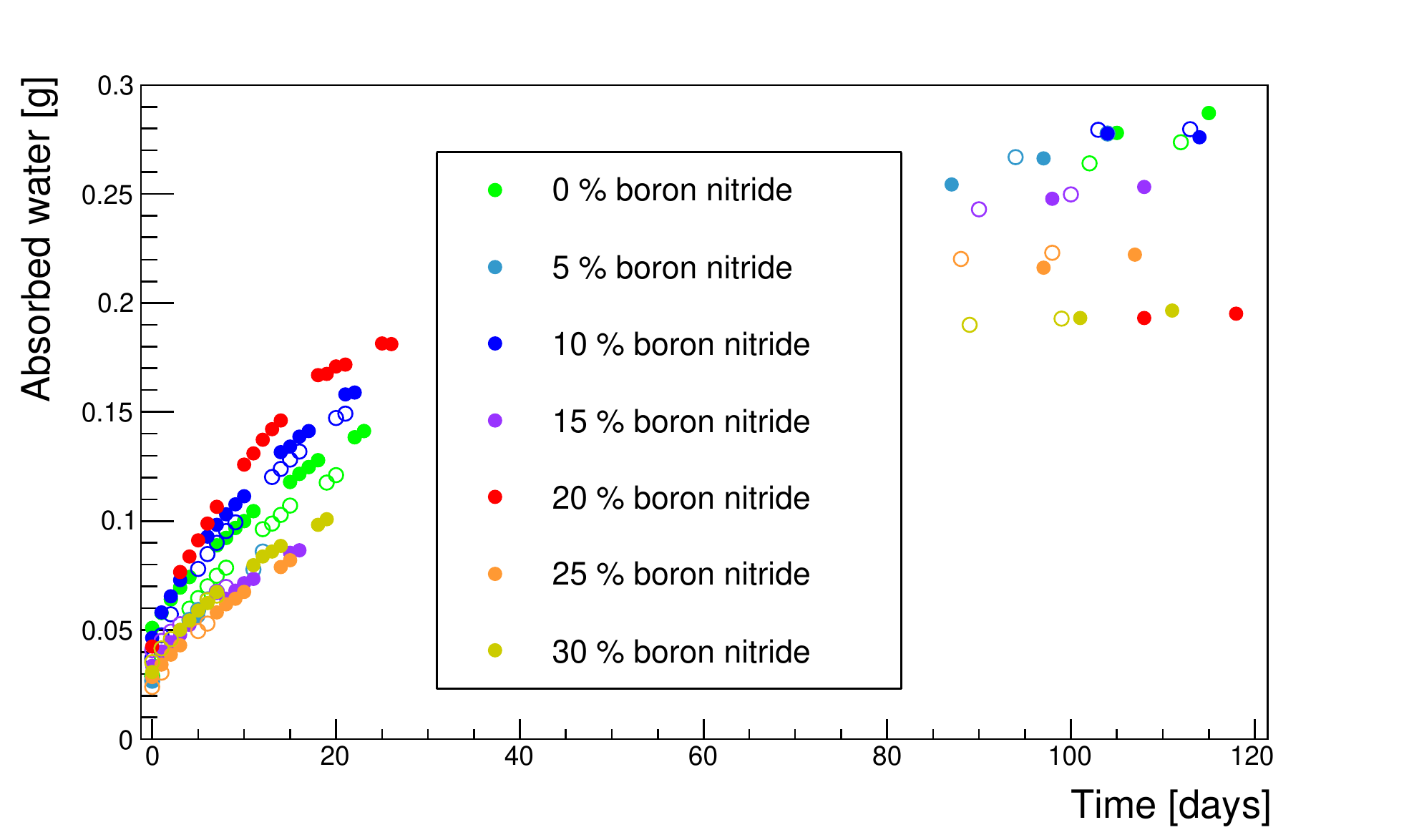}
\caption{Total absorbed water per glue sample over time: oven cured samples ({\unit[0, 10 and 20]{\%}} boron nitride) showed a faster initial absorption than the other samples, presumably due to the higher air content. The final amount of absorbed moisture was lower for samples with higher filling content, since the moisture content absorbed by boron nitride is negligible~\cite{BN}. Due to limited access, measurements were paused between days 35 to 90.}
\label{fig:soaking}
\end{figure}

After performing the measurements, all samples were stored in a closed box next to an open water source in order to quantify the maximum amount of absorbed water (``soaking''). All samples were weighed daily to track their moisture content increase. Figure~\ref{fig:soaking} shows the increase in absorbed moisture for all glue samples over time.
The humidity absorption was found to follow the expected behaviour showing fast absorption initially and slow saturation in the long term~\cite{absorption}.

The obtained data was used to calculate the absorbed humidity content at saturation as well as the moisture content the samples had after curing (see table~\ref{tab:moisture}).
\begin{table}
 \centering
 \begin{tabular}{c|c|c|r|r|r|r}
 & & & \multicolumn{4}{l}{Moisture content, $[$\%$]$} \\
BN content & Sample & Dry weight & \multicolumn{2}{c}{After curing} & \multicolumn{2}{c}{After soaking} \\
$[$\%$]$ & & $[$g$]$ & & -BN & & -BN \\
\hline
0 & 1 & 3.10 & 0.9 & - & 9.3 & - \\
0 & 2 & 3.12 & 1.1 & - & 8.8 & - \\
5 & 1 & 2.87 & 0.8 & 0.9 & 9.3 & 9.8 \\
5 & 2 & 2.93 & 0.8 & 0.9 & 9.5 & 10.0 \\
10 & 1 & 2.95 &	0.5 & 0.6 & 9.4 & 10.5 \\
10 & 2 & 3.00 & 0.6 & 0.6 & 9.3 & 10.3 \\
15 & 1 & 3.03 & 0.7 & 0.9 & 8.4 & 9.8 \\
15 & 2 & 3.06 & 0.8 & 0.9 & 8.2 & 9.6 \\
20 & 1 & 2.22 & 0.6 & 0.7 & 8.8 & 11.0 \\
20 & 2 & 2.91 & 0.7 & 0.8 & 8.7 & 10.9 \\
25 & 1 & 3.00 & 0.7 & 0.9 & 7.4 & 9.9 \\
25 & 2 & 2.98 & 0.7 & 0.9 & 7.5 & 10.0 \\
30 & 1 & 2.93 & 0.8 & 1.2 & 6.7 & 9.6 \\
30 & 2 & 2.91 & 0.8 & 1.2 & 6.6 & 9.5 \\
\end{tabular}
\caption{Moisture content of glue samples with different filling contents after curing and soaking, in weight percent compared to baked samples (dry weight). Since boron nitride absorbs negligible amounts of water, samples with higher filling content absorb less moisture than sample with lower filling percentage. Correcting for the boron nitride content (columns ``-BN''), all samples show similar absorption behaviour.}
\label{tab:moisture}
\end{table}

Comparing the different glue samples showed that, when correcting for the boron nitride filling content, all samples absorb moisture to a similar degree. After curing, samples had a humidity content of approximately \unit[1]{\%} of their overall glue weight (excluding the boron nitride filling). At saturation, their humidity content was about \unit[10]{\%} of the overall same glue weight. Oven cured samples with boron nitride were found to absorb slightly more moisture (\unit[10.3 to 11.0]{\%}) than room temperature cured samples (\unit[9.3 to 10.0]{\%}), presumably due to their higher content of entrapped air.

Samples after preparation could therefore be estimated to contain about \unit[10]{\%} of their maximum moisture content at saturation (individual values varied between \unit[6 and 12]{\%}, presumably due to different levels of humidity during sample preparation combined with varying amounts of time between sample preparation and first weighing).

\section{Performed measurements}
\label{sec:imaging}

The coefficient of moisture expansion $\alpha_{CME}$ is defined as
\begin{equation}
 \alpha_{CME} = \frac{\Delta l/l}{m_{water}/m_{sample}},
\end{equation}
where $\Delta l/l$ is the relative length change referred to as strain, $m_{water}$ is the weight of the absorbed water and $m_{sample}$ is the dry weight of the sample.

\begin{figure}[htp]
\centering
\includegraphics[width=0.5\linewidth]{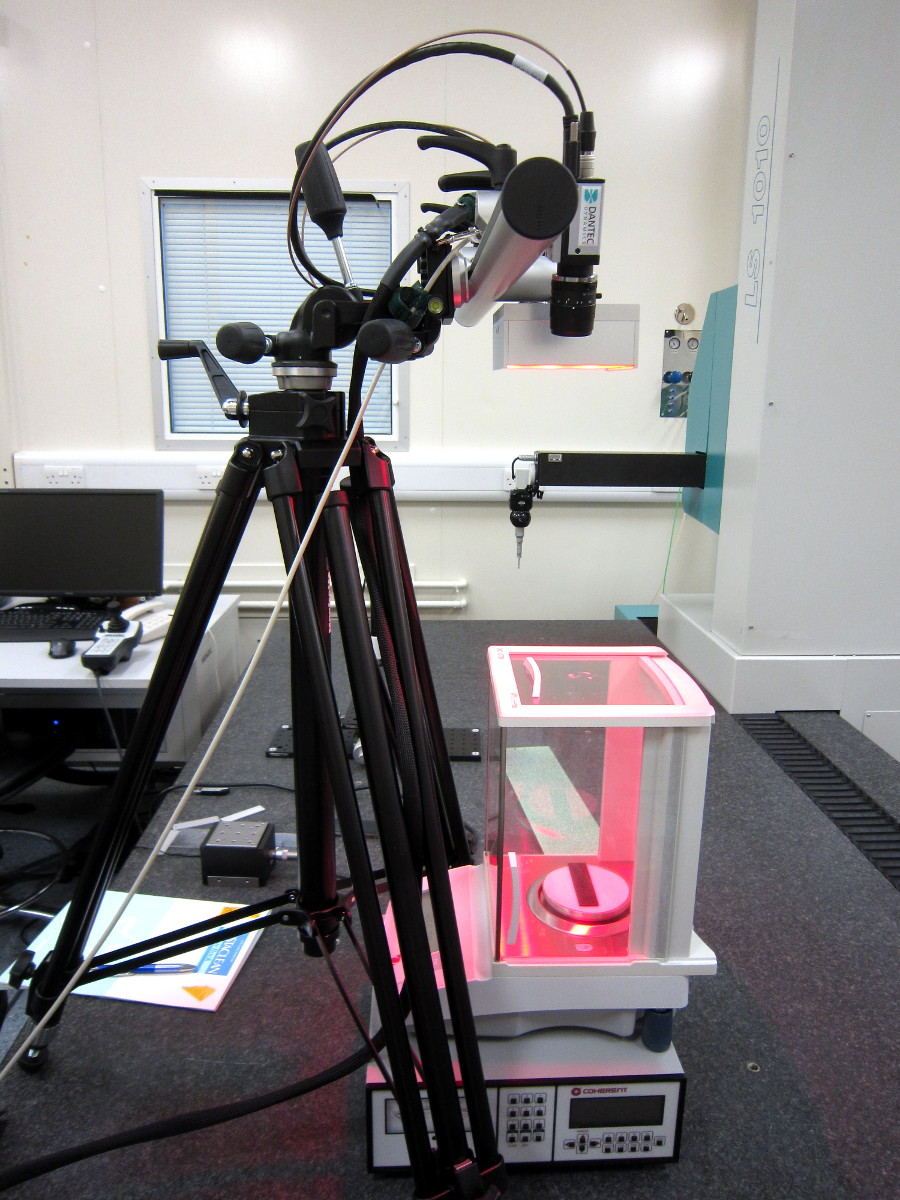}
\caption{Measurement setup: two cameras mounted on a tripod were pointed at a sample positioned on scales: the weight of the sample was monitored using the scales and after defined steps of weight increase, a set of images was taken of the sample.}
\label{fig:setup}
\end{figure}

The sample expansion was measured using a DANTEC Dynamics DIC Standard 3D image correlation system, set up according to good practice recommendations for the samples under investigation~\cite{IMCOR}.
Two cameras mounted on a tripod were pointed at a sample under investigation (see figure~\ref{fig:setup}) and images from both cameras were overlayed to compensate for any angular distortion. Since the samples under investigation expanded mainly in-plane with limited strain occurring out-of-plane, the cameras were arranged at a small angle of about \unit[20]{$^{\circ}$} between them. Prior to a measurement, the cameras were warmed up and calibrated under the same lighting conditions as used throughout the measurement.
The associated, commercial software (Istra4D) was used to compare all images to an initial reference picture: at the beginning of each measurement, one reference image was taken. Subsequent images captured after water had been absorbed were compared to the initial reference image, which calculated expansion or contraction with respect to the reference sample based on position changes of the sample speckle pattern.

Since the process of spray painting occasionally led to areas with higher or lower concentrations of speckles (see figure~\ref{fig:speckles}), the uniformity of the reconstructed strain was verified as part of each measurement: in addition to calculating the strain over the full sample surface, smaller sample areas were selected for a comparative reconstruction. If the strain reconstructed in a given area deviated too much from the strain determined using the full sample area, the sample's speckle pattern was removed and re-applied.

Calculating the occuring expansion from an image with respect to the original reference image required the sample under investigation to remain in the same position. Therefore, samples were placed on scales during the measurement (determining the sample weight down to $\mathcal{O}(\unit[0.1]{\text{mg}})$), see figure~\ref{fig:setup}, with the full sample within the field of view of both cameras for the entire measurement. Samples were placed on a thin sheet of paper to provide the recommended neutral, high-contrast background. The measurement was set up in a cleanroom environment for stable temperature and humidity both to prevent environmental changes from affecting the measured strain and to avoid surface contaminations from impacting the imaging process.

A measurement was conducted by placing each dried out sample on the scales next to an open water container and monitoring the total sample weight. Every time the sample weight had increased by \unit[0.4-0.5]{mg}, the image system was triggered and five images of the sample were taken through the opened roof of the scales. In order to maintain alignment with the original sample orientation, the sample was not moved throughout the measurement: the scales' top closure was opened for image recording and closed again for a better weight measurement accuracy. For each sample, the humidity increase was monitored over the course of several hours to include at least \unit[10]{mg} of absorbed water.

\begin{figure}[htp]
\centering
\includegraphics[width=0.8\linewidth]{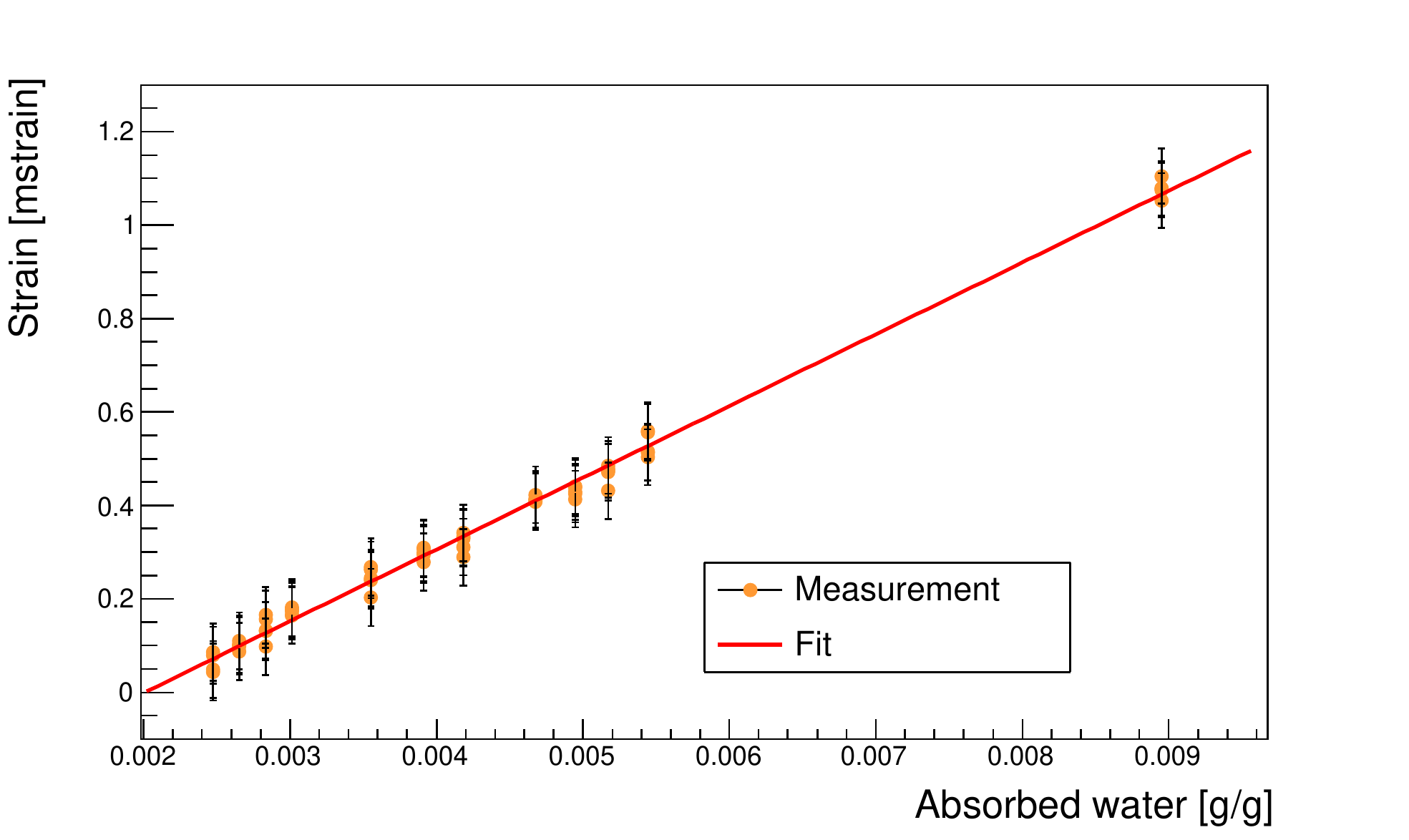}
\caption{Absorbed water content and corresponding strain for a glue sample with {\unit[20]{\%}} boron nitride content. Strains were measured in millistrain (with {\unit[1]{mstrain}} corresponding to an extension of {\unit[1]{mm}} per {\unit[1]{m}}). A linear fit is performed to determine the sample's CME.}
\label{fig:plotfit}
\end{figure}

In order to calculate the overall CME, the measured strain was related to the absorbed moisture content (calculated from the measured weight divided by the dry mass of the glue sample) and a linear fit function applied (see figure~\ref{fig:plotfit}).

\section{Results}

\begin{figure}[htp]
\centering
\includegraphics[width=0.8\linewidth]{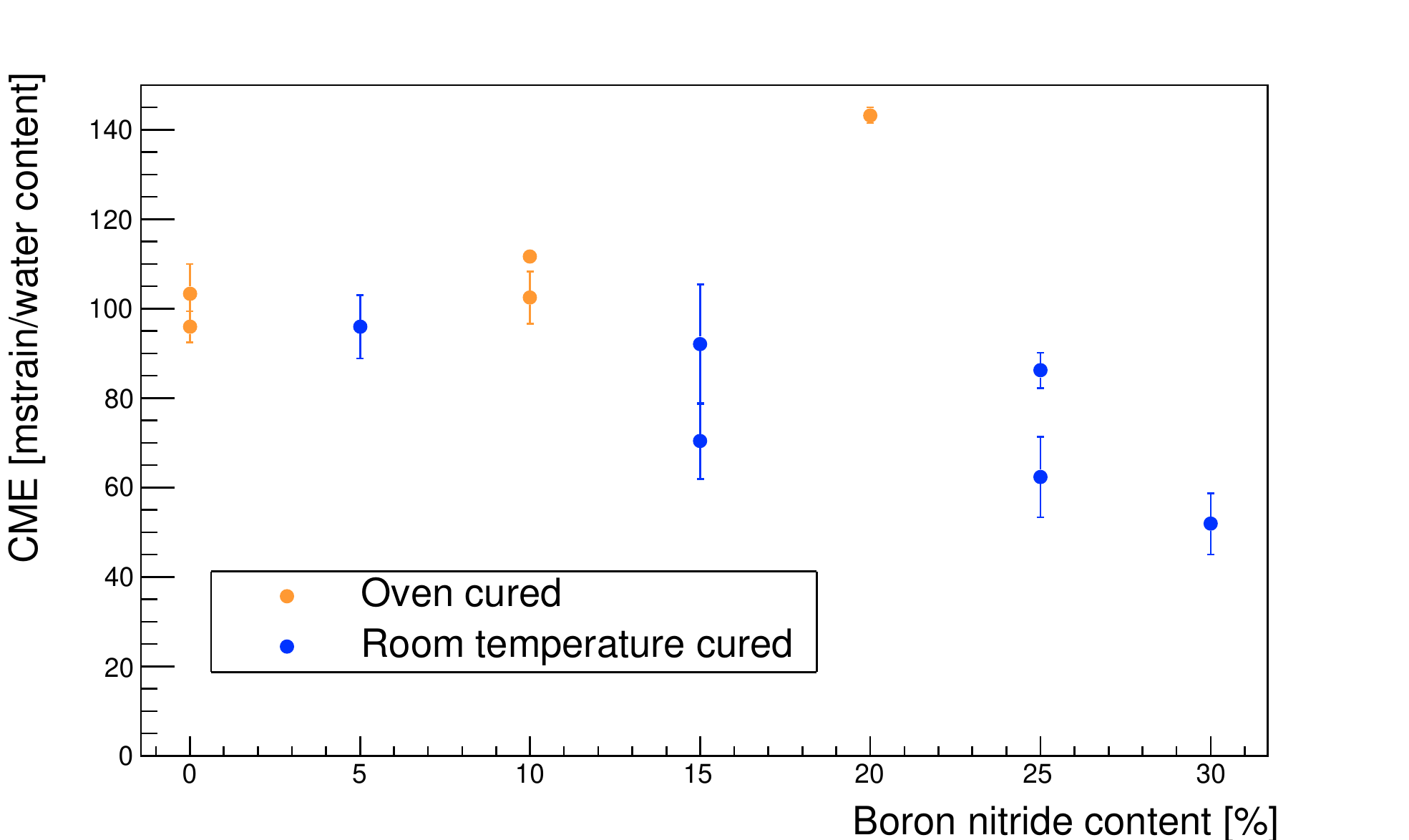}
\caption{CME measured for samples of Hysol with different percentages of boron nitride powder. The measured values correspond to two samples prepared for each percentage of boron nitride content, where available. The displayed error bars correspond to the uncertainties determined in the linear parameter fit performed using ROOT~\cite{ROOT}.}
\label{fig:results}
\end{figure}

The determined CME for all samples under investigation are shown in Figure~\ref{fig:results}.

\begin{figure}[htp]
\centering
\includegraphics[width=0.8\linewidth]{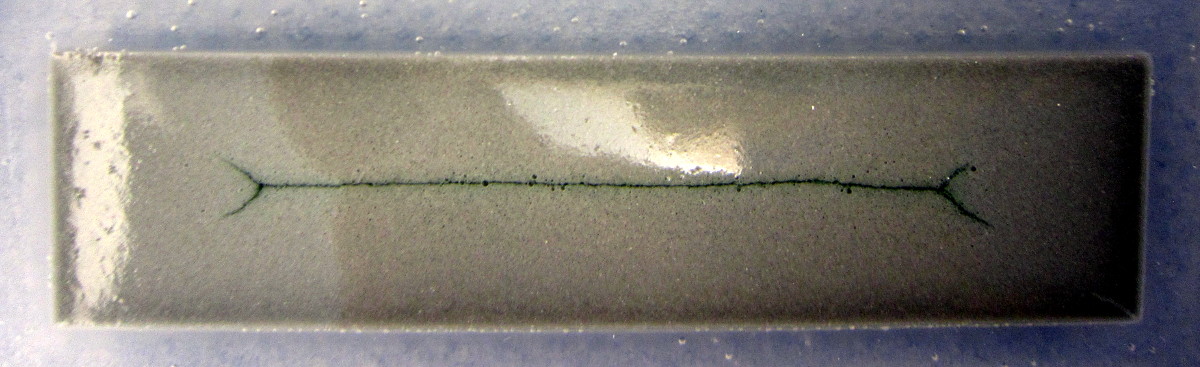}
\caption{Top surface of a Hysol sample with a boron nitride content of {\unit[5]{\%}}. During the curing process, the movement of air through the glue volume caused a partial separation of adhesive and filling.}
\label{fig:uneven}
\end{figure}

For samples cured at room temperature, a higher boron nitride content resulted in a lower CME as expected. For oven cured samples loaded with boron nitride, the high air content (see figure~\ref{fig:bubbles}) led to an overall higher CME. Additionally, room temperature cured samples were found to show a larger discrepancy between samples with the same boron nitride content depending on which side had been milled off. This effect matches an observation made for room temperature cured samples, where in the process of curing, air moving through the glue volume led to a movement of the mixed glue, which caused separation of adhesive and boron nitride filling (see figure~\ref{fig:uneven}). As a result, different regions of the glue sample can be assumed to have different compositions, which result in larger discrepancies than for oven cured samples, where the shorter curing time prevented a similar extent of separation.

The CME of Hysol 9396 with boron nitride powder filling was measured for filling percentages between \unit[0 and 30]{\%}. It was found to range from \unit[50 to 100]{mstrain/(g$_{\text{water}}$/g$_{\text{adhesive}}$)} for room temperature cured samples and to reach up to \unit[140]{mstrain/(g$_{\text{water}}$/g$_{\text{adhesive}}$)} for an oven cured sample with high boron nitride content (\unit[20]{\%}) that had not been de-gassed after mixing and served as a worst case example.

In order to compare the measured CME with the known CTE of the adhesive under investigation (see table~\ref{tab:CMECTE}), the following assumptions were made:
\begin{itemize}
 \item samples absorbed about \unit[1]{\%} of their adhesive weight under normal storage conditions (see section~\ref{sec:absorb}, table~\ref{tab:moisture})
 \item during operation in the ATLAS detector, the adhesive would experience a temperature drop of \unit[-60]{$^{\circ}$C} (from assembly at room temperature to an operating temperature at~\unit[-40]{$^{\circ}$C}~\cite{TDRs})
 \item the CTE of glue samples with boron nitride filling is approximately the CTE of the adhesive multiplied by the adhesive content, since the CTE of boron nitride is negligible compared to the CTE of Hysol 9396~\cite{BNCTE}
\end{itemize}
The CTE for Hysol 9396 was obtained from the datasheet: \unit[70.7]{$\upmu$m/m/$^{\circ}$C}~\cite{Hysol}.

\begin{table}
 \centering
 \begin{tabular}{r|r|c|r|c}
  & CME & Strain & CTE & Strain \\
Sample & measured & from \unit[1]{\%} MC & calculated & from \unit[-60]{$^{\circ}$C} \\
  & $[$mstrain/(g/g)$]$ & $[$mstrain$]$ & $[\upmu$strain/$^{\circ}$C$]$ & $[$mstrain$]$ \\
  \hline
\unit[0]{\%} I & $95 \pm 7$ & $0.95 \pm 0.07$ & 71 & 4.24 \\
\unit[0]{\%} II & $103 \pm 13$ & $1.03 \pm 0.13$ & 71 & 4.24 \\
\unit[5]{\%} I & $98 \pm 15$ & $0.93 \pm 0.14$ & 67 & 4.03 \\
\unit[10]{\%} I & $112 \pm 3$ & $1.01 \pm 0.03$ & 64 & 3.82 \\
\unit[10]{\%} II & $103 \pm 12$ & $0.93 \pm 0.11$ & 64 & 3.82 \\
\unit[15]{\%} I & $71 \pm 17$ & $0.60 \pm 0.14$ & 60 & 3.61 \\
\unit[15]{\%} II & $97 \pm 28$ & $0.81 \pm 0.24$ & 60 & 3.61 \\
\unit[20]{\%} I & $140 \pm 3$ & $1.14 \pm 0.02$ & 57 & 3.39 \\
\unit[25]{\%} I & $86 \pm 8$ & $0.65 \pm 0.06$ & 53 & 3.18 \\
\unit[25]{\%} II & $61 \pm 19$ & $0.46 \pm 0.14$ & 53 & 3.18 \\
\unit[30]{\%} I & $52 \pm 14$ & $0.36 \pm 0.10$ & 49 & 2.97 \\
\end{tabular}
\caption{CME and CTE in comparison for Hysol samples with different percentages of boron nitride filling. In addition to the measured CME, the total strain expected from drying a sample with a realistic moisture content of {\unit[1]{\%}} is provided (column ``from {\unit[1]{\%}} MC'') for a comparison with the corresponding expected strain from cooling the same glue by {\unit[-60]{$^{\circ}$C}} (column "from {\unit[-60]{$^{\circ}$C}}``).}
\label{tab:CMECTE}
\end{table}

\section{Conclusion}

Measurements were performed to study the moisture absorption of Hysol 9396 samples with different filling percentages of boron nitride powder. It was found that all samples absorbed approximately \unit[10]{\%} of their adhesive weight in moisture until saturation. Samples had absorbed approximately \unit[1]{\%} of their adhesive weight in moisture after curing, independent of the filler content.

In the absence of degassing, the addition of boron nitride powder to a Hysol mix was found to impact the glue homogeneity: oven cured samples showed a high number of air bubbles trapped inside the glue volume during the curing process. Samples with a comparable filler content cured at room temperature showed a lower air content, but the movement of air escaping from the sample during curing was found to lead to a larger degree of filler separation and an inhomogeneous filler distribution in the sample. 

The CME of all samples under investigation was measured by simultaneously monitoring the gained weight from moisture absorption and the corresponding strain with an image correlation system. The determined CME values varied between \unit[50 and 140]{mstrain/(g/g)}, depending on boron nitride content and curing method.

A comparison of the hygrometric strain corresponding to samples with \unit[1]{\%} moisture content drying and the thermoelastic strain accompanying a cooling by \unit[60]{$^{\circ}$C} was performed. The strain associated with drying was found to be smaller but reaching up to \unit[30]{\%} of the strain associated with cooling. Accordingly, glue samples with a moisture content of more than \unit[1]{\%} would lead to larger strains occurring during drying.

The measurements therefore confirmed that for an accurate description of the strains developing in ITk support structures during operation, the CME of the used adhesives should be included in addition to its CTE.

\section*{Acknowledgments}

The work presented here was partially funded by the Canada Foundation for Innovation (CFI) and the Natural Sciences and Engineering Research Council (NSERC) of Canada.

\bibliographystyle{unsrt}
\bibliography{bibliography.bib}

\end{document}